\documentclass[aps,prl,showpacs,groupedaddress,superscriptaddress,twocolumn]{revtex4}

\usepackage{amsmath,amsfonts,amssymb,graphics,graphicx,epsfig,color,bbm,natbib,bm,times}

\begin{document}

\title{Local and Global Distinguishability in Quantum Interferometry}

\author{Gabriel A.\ Durkin} \email{gabriel.durkin@qubit.org}
\affiliation{Jet Propulsion Laboratory, California Institute of Technology, Pasadena, California 91109, USA}

\author{Jonathan P. Dowling} 
\affiliation{Hearne Institute for Theoretical Physics, Louisiana State University, Baton Rouge, Louisiana 70803, USA }

\date{19th December 2006}


\begin{abstract} A statistical distinguishability based on relative entropy characterises the fitness of quantum states for phase estimation. This criterion is employed in the context of a Mach-Zehnder interferometer and used to interpolate between two regimes, of local and global phase distinguishability. The scaling of distinguishability in these regimes with photon number is explored for various quantum states. It emerges that local distinguishability is dependent on a discrepancy between quantum and classical rotational energy. Our analysis demonstrates that the Heisenberg limit is the true upper limit for local phase sensitivity. Only the `NOON' states share this bound, but other states exhibit a better trade-off when comparing local and global phase regimes. 
\end{abstract}

\maketitle

Interferometry may be viewed as estimation of a finite phase parameter from a position of prior ignorance \cite{Phase-est-algorithm}. It is also used to identify small changes in a known phase, or to track such changes over time. The tasks of \emph{global} phase acquisition and \emph{local} phase tracking are both important challenges. Classically, the distinction is well-understood, for example in implementations of Radar/Sonar \cite{Radar-Local/Global-Phase-Est}. Any comprehensive analysis of a \emph{quantum} interferometer should address these two different facets of metrology \cite{Qm-Metrology}. 

Real interferometers always have trade-offs between performance (e.g. accuracy and precision), robustness (to photon loss and decoherence), and complexity (resources in state generation, phase encoding and measurement).  Focusing on the performance aspect of an \emph{ideal} interferometer free of losses and decoherence may clarify which quantum correlations lead to precision enhancement \cite{Heisenberg-Limit,Yurke-SU2-Interfer} in local and global limits. In this paper we examine the intrinsic fitness of various quantum states for interferometry independent of any specific estimation protocol, i.e. irrespective of how the measurement data are processed. The fitness criteria introduced will be based on the collective information content of the measurement distribution, and not simply on features like the mean or variance.

A quantum state of $n$ photons distributed across two spatial modes $a$ and $b$ is isomorphic to a spin-$j$ particle. A two-mode Fock state $|n_a\rangle_a\otimes|n_b\rangle_b$ is mapped onto $|j,m\rangle_z$ where $j = (n_a + n_b)/2$ and subscript $z$ denotes that this is an eigenstate of $\hat{J}_z$ with eigenvalue  $m = (n_a - n_b)/2$. In terms of creation and annihilation operators, generators of unitary transformations representing linear optical elements are: $\hat{J}_{x}=   \frac{1}{2}  (\hat{a}^{\dagger} \hat{b}+ \hat{b}^{\dagger} \hat{a} )$, $ \hat{J}_{y}= -\frac{i}{2} (\hat{a}^{\dagger} \hat{b} - \hat{b}^{\dagger} \hat{a} )$ and $\hat{J}_{z}=\; \;  \frac{1}{2} (\hat{a}^{\dagger} \hat{a} - \hat{b}^{\dagger} \hat{b})$. Also, $\hat{J}^2 = \hat{n}/2(\hat{n}/2+1)$ where $\hat{n} = (\hat{a}^{\dagger} \hat{a} + \hat{b}^{\dagger} \hat{b})$ is the total photon number. The generators obey commutation relations, $[\hat{J}_{\alpha},\hat{J}_{\beta}]=2i \epsilon_{\alpha \beta \gamma} J_{\gamma}$ and the eigen-equations are $\hat{J}^2 | j,m \rangle_i = j(j+1)| j,m \rangle_i$ and  $\hat{J}_i | j,m \rangle_i = m | j,m \rangle_i$ for $i \in \{x,y,z \}$. 

\begin{figure} 
\includegraphics[width=1.6in]{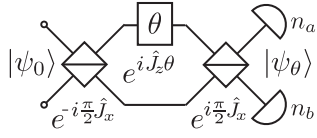}
\caption{\small{Mach-Zehnder interferometer with $2$-mode input state $|\psi_0 \rangle$. Angle $\theta$ is the relative phase between the two arms. Beamsplitters perform $\pm \frac{\pi}{2}$ rotations about the $x$ axis. Detectors at the outputs count photon numbers $n_a$ and $n_b$. }} \label{MZ}
\end{figure}

Given an input or probe state $| \psi_{0} \rangle$, the Mach-Zehnder (MZ) interferometer with phase difference $\theta$ between the arms (FIG.\ref{MZ}) performs a transformation $\exp\{i \hat{J}_{x} \pi /2 \} \exp\{i \hat{J}_{z} \theta \} \exp\{-i \hat{J}_{x} \pi /2 \}$, equivalent to a rotation  \cite{Yurke-SU2-Interfer} through $\theta$ about the $y$-axis, i.e. $\exp\{i \hat{J}_{y} \theta\}$. One infers an unknown $\theta$ by making measurements on the output state $| \psi_{\theta} \rangle$. This task is hampered by the non-existence of a Hermitian phase operator $\hat{\theta}$ in Hilbert space. Even in this idealised lossless setting, any estimate $\phi_e$ of the phase has accuracy and precision limited by both the choice of $| \psi_{0} \rangle$ and the measurement employed. Conventionally and most simply one counts photons in the output arms of the interferometer, equivalent to measurement of $\hat{J}_z$, and this will be considered here. Other measurements have been proposed, e.g. $\hat{J}^{2}_z$ \cite{HLimited-Interfer-Decorr}, parity measurements \cite{parity-measurements}, homodyne detection  \cite{Shapiro-Homodyne,Hradil-Quad}, heterodyne detection \cite{Shapiro-Heterodyne} and forms of generalised measurement \cite{Qm-Metrology,Phase-POVM, Fisher-Optimal-measurements}. 

The evolution of a pure state in an ideal MZ interferometer is governed by the Schr\"{o}dinger equation,
\begin{equation} \label{Schr}
i \frac{\partial}{\partial \theta} | \psi_{\theta} \rangle = \hat{J}_{y} | \psi_{\theta} \rangle
\end{equation}
for $\theta$ a time-like variable and $\hbar =1$. The spin observable $\hat{J}_y$ plays the role of Hamiltonian and as such is a conserved quantity, its eigenvectors are preserved. The transformed state $| \psi_{\theta} \rangle$ is
\begin{equation}\label{evo}
 e^{i \hat{J}_{y} \theta} | \psi_{0} \rangle \!   = \! \! \! \sum_{m=-j}^{+j} \! \! |j,m\rangle \!_z  \: _z \! \langle j,m | \psi_{\theta} \rangle 
=\! \! \! \sum_{m=-j}^{+j} \psi_m(\theta) |j,m\rangle_z 
\end{equation}
in the $\hat{J}_z$ basis. The probability amplitudes $ \psi_m(\theta)$ are $r_{m}(\theta) \exp\{i \phi_m (\theta)\} $ in polar form, the modulus $r_m$ and argument $\phi_m$ both real-valued functions of $\theta$. The measurement distribution is the set $P(\theta) = \{p_m(\theta)\}$, where $p_m(\theta) = \left| \: _z \! \langle j , m  | \psi_{\theta} \rangle \right|^2 = \psi_m^* (\theta) \psi_m (\theta) = r_m^2 (\theta)$. These probabilities lie at the heart of phase estimation -- over a data run the frequencies of measurement outcomes tend towards the same distribution. The full distribution  $\{p_m(\theta)\}$ carries information about $\theta$, more than is characterized by its maxima \cite{MLE-vs-Bayes}, or that is contained in means and variances, or particular moments of $\hat{J}_z$. This is an important detail when the governing probabilities $p_m(\theta)$ are multiple-peaked and non-gaussian in $\theta$, as is often the case in quantum interferometry \cite{Smerzi-MZ}.

By matching ratios of observed measurements to a particular governing distribution one may infer $\theta$ to within some precision. However, two issues prevent this from being done exactly and unambiguously. Firstly, the number of experimental trials is finite, and may be small, so the set of measured outcomes may not be typical of the governing distribution. For example, a fair coin tossed four times may still land `heads' four times in a row, an atypical result. Secondly, even for typical data sets there may not exist a one-to-one mapping $\theta \leftrightarrow P(\theta)$  between phases and probability distributions for $\theta \in [0,2\pi)$. Without such a mapping there is no simple inversion, $P(\theta) \mapsto \theta$.

What properties of $p_m(\theta)$ might make it more immune to these difficulties? Ideally, one asks that $p_m(\theta_1) \neq p_m(\theta_2)$ and that these distributions are somehow `far apart', in the sense that one would be unlikely to mistake one for the other. From the Theory of Types \cite{Cover and Thomas}, the probability $p_{(2 \rightarrow 1)}$ that a parent distribution $P(\theta_2)$ gives rise to a data set typical of $P(\theta_1)$ is bounded by:
\begin{equation}\label{prob-mistake}
\frac{2^{-k S[P(\theta_1) || P(\theta_2)]} }{(k+1)^{(2j+1)}} \leq p_{(2 \rightarrow 1)} \leq 2^{-k S[P(\theta_1) || P(\theta_2)]}
\end{equation}
for a sequence of $k$ independent measurements. Here $(2j+1)$ is the cardinality of distinct measurement outcomes $m \in \{-j,-j+1,\dots,+j\}$. The non-negative functional $S[P(\theta_1) || P(\theta_2)]$ is known as the relative entropy or Kullback-Leibler divergence \cite{Kullback-Leibler},
\begin{equation}
S[P(\theta_1) || P(\theta_2)] = \sum_{m =-j}^{+j} p_{m}(\theta_1) \log_2 \frac{ p_{m}(\theta_1)}{ p_{m}(\theta_2)} \; ,
\end{equation}
quantifying the distinguishability of one distribution from another. If better-than-even odds were required in `distinguishing' distribution $P_2$ from its neighbor $P_1$ after one measurement, Eq.\eqref{prob-mistake} imposes a lower bound $S[P_1||P_2] > 1$. Relative entropy has been employed before \cite{Hradil-Quad} in the context of a `maximum likelihood' approach to phase estimation, but now we use it to find the \emph{intrinsic} fitness of states for this task, without reference to a particular estimation protocol. 

We propose that a global \emph{distinguishability} for estimating a phase $\theta$, previously known to exist within a finite interval $\Delta$ centred on $\theta = \chi$ be defined as:
\begin{equation}
\mathcal{D}(\chi, \Delta) \! = \! \frac{1}{\Delta^2} \int_{\theta_1, \theta_2 = \chi - \Delta/2}^{ \chi + \Delta/2} \! \! \! \!  S[P(\theta_1) || P(\theta_2)] d \theta_1  d \theta_2 .
\end{equation}
This is the arithmetic mean of relative entropies between all pairs of distributions originating within this phase interval. In terms of the upper bounds to probabilities in Eq.\eqref{prob-mistake}, $2^{-k \mathcal{D}}$ is a geometric mean. Note also that while $S(P_1||P_2) \neq S(P_2||P_1)$, quantity $\mathcal{D}$ is symmetric; it contains the sum $S(P_1||P_2) + S(P_2||P_1)$ for all probability pairs.

Parameter $\Delta$ may be considered an \emph{a priori precision}, and $\chi$ an \emph{a priori estimate}; the mean phase across the $\Delta$ interval for a uniform prior probability distribution. The size of $\Delta$ has a pronounced effect on the properties of $\mathcal{D}$. Since $\theta$ is a cyclic variable, $p_{m}(\theta) = p_{m}(\theta + 2 \pi)$ and  $\Delta < 2 \pi$ is a necessary prerequisite to performing the inversion $\{p_{m}(\theta) \} \mapsto \theta$. In addition to translation symmetries there may exist mirror symmetries (depending on the input state) and distinguishability will be lower for $\Delta$ intervals inclusive of these symmetry points. A strength of our analysis is the freedom to operate beyond a restricted neighborhood like $\theta \approx 0$, cf. Refs.\cite{parity-measurements,Smerzi-MZ,Hradil-Fisher}.

One may confine $\theta$ to an interval $\Delta \ll 2 \pi$, given adequate prior phase knowledge. In this case it is quite possible for a phase uncertainty to be greater after the measurement than before. This is because the direct measurement of $J_z$ is unsharp -- such a discretely-valued measurement can only impart partial information about the continuous phase parameter $\theta$. For vanishingly small $\Delta \sim 0$, then $\mathcal{D}$ approaches a \emph{local} distinguishability:
\begin{align}& \! \! \left. \left. \frac{1}{4}  \right[ \! \!  S[P(\chi \! - \! \! \Delta/2)||P(\chi \! + \! \! \Delta/2)] \! +   \!  S[P(\chi \! + \! \! \Delta/2)||P(\chi \! - \! \! \Delta/2)] \right] \! \nonumber \\
& = \!  \frac{\Delta^2}{8 \ln 2} \left\{ \: \mathcal{J}(\chi\! - \! \Delta/2) + \mathcal{J}(\chi\! + \! \Delta/2) \: \right\} + O(\Delta^3)\; , \label{Fishercomesfrom}
\end{align}
Above, $\mathcal{J}(\theta)$ is the classical Fisher Information \cite{Cramer-Fisher} for the measurement distribution $P(\theta)$,
\begin{equation}
\mathcal{J}(\theta) = \sum_{m =-j}^{+j}p_m (\theta)\left(  \frac{\partial }{\partial \theta} \ln p_m (\theta) \right)^2 \; . \label{F1}
\end{equation}
The lower line of Eq.\eqref{Fishercomesfrom} comes from a series expansion of $S[P(\chi \pm \Delta/2)||P(\chi \mp \Delta/2)]$ about $(\chi \pm \Delta/2)$ to second order in $\Delta$. The Fisher information gives a measure of the information contained in the \emph{full} distribution $P(\theta)$ about the parameter $\theta$ \cite{Cover and Thomas}. It is also the unique distance metric on the manifold of probability distributions \cite{Fisher-unique-metric}. 
With the knowledge that $\mathcal{J}(\theta)$ provides the scaling factor of local distinguishability, we now derive an explicit form in terms of the Hamiltonian and the measurements. Taking the definition, Eq.\eqref{F1} and substituting $p_m(\theta)= r_m^2(\theta)$,
\begin{equation}\label{F2}
\mathcal{J}(\theta) = 4 \sum_{m =-j}^{+j} \dot{r}_m^2(\theta) \; ,
\end{equation}
where derivatives with respect to $\theta$ are denoted by an overdot. If instead one substitutes $p_m(\theta) = \psi^*_m(\theta) \psi_m(\theta)$,
\begin{equation} \label{secondF}
\mathcal{J}(\theta) = \sum_{m =-j}^{+j} \frac{ \psi_m}{\psi^*_m} \dot{\psi}^{* 2}_m + \frac{\psi^*_m}{\psi_m} \dot{\psi}^{2}_m + 2 \dot{\psi}^{*}_m \dot{\psi}_m  \; .
\end{equation}
The last term may be evaluated using Eq.\eqref{Schr}, $\sum_{m}  \dot{\psi}^{*}_m \dot{\psi}_m  = \langle \dot{\psi}_{\theta}|\dot{\psi}_{\theta}\rangle = -i\langle \psi_{\theta}|\hat{J}_y^{\dagger} \times i \hat{J}_y |\psi_{\theta} \rangle = \langle \hat{J}_{y}^{2} \rangle$. The other terms under the summation of Eq.\eqref{secondF} can be simplified by putting $\psi_m = r_m \exp\{i \phi_m \}$,
\begin{equation} \label{F3}
\mathcal{J}(\theta) =  \sum_{m =-j}^{+j} 2 \dot{r}_m^2 - 2 r_m^2 \dot{\phi}_m^2 + 2 \langle \hat{J}_{y}^{2} \rangle \; .
\end{equation}
Comparing Eq.\eqref{F3} and Eq.\eqref{F2} gives
\begin{equation} \label{Ffinal}
\mathcal{J}(\theta) /4 =  \langle \hat{J}_{y}^{2} \rangle  - \sum_{m =-j}^{+j} r_m^2 \dot{\phi}_m^2  \;  =  \; \langle \hat{J}_{y}^{2} \rangle  - \langle \dot{\Phi}^2 \rangle_c \; 
\end{equation}
The final term with subscript $c$ denotes a classical mean with respect to distribution $\{p_m(\theta)\}$, for the square of a random variable $\dot{\Phi}$ taking values $\dot{\phi}_m$. In analogy with Ref.\cite{Hall-Kinetic-Fisher} the Fisher information for any $| \psi_{\theta} \rangle$ is exactly the \emph{discrepancy} between  quantum rotational energy and the rotational energy associated with $2j+1$ classical point objects, one for each $m$ value. Note that Eq.\eqref{Ffinal} holds in general for any hermitian Hamiltonian $\hat{J}_y \mapsto \hat{H}$ and for spin measurements $m$ made along \emph{any} direction in Euclidean space, $z \mapsto z'$.

From Eq.\eqref{Ffinal} we may develop a type of uncertainty relation connecting $\theta$, the Hamiltonian and the measurement basis as follows. An explicit lower bound on the mean-squared error of an unbiased \cite{unbiased} estimate $\phi_e$ on the true phase $\theta$ is given by the reciprocal of the Fisher information:
\begin{equation}
(\delta \phi_e)^2 \geq 1 / \mathcal{J}(\theta) \; ,
\end{equation}
called the Cram\'{e}r-Rao bound \cite{Cramer-Fisher}. Therefore, for optimal $\theta$ precision one maximizes the Fisher information \cite{Braunstein-knee}. The bound is well-known in information theory, having a general applicability not shared by a popular linearised error model, $\: \delta \phi_e \:  \left|\partial \langle  \hat{J}_z \rangle / \partial \theta \right| \approx [\langle \hat{J}_z^{2} \rangle - \langle \hat{J}_z \rangle^{2}]^{1/2}$, \cite{Qm-Metrology,Yurke-SU2-Interfer,HLimited-Interfer-Decorr,parity-measurements} which produces incorrect or inconclusive scaling of $ \delta \phi_e $ for certain states $|\psi_\theta \rangle$ \cite{Smerzi-MZ,Hradil-Fisher}. The Cram\'{e}r-Rao bound provides a proof of the `Heisenberg' precision limit \cite{Qm-Metrology,Heisenberg-Limit} as follows: For certain input states $ \dot{\phi}_m = 0$  and $\langle \hat{J}_{y}^{2} \rangle$ is bounded from above by its largest eigenvalue, i.e. $j^2$. Therefore for a state of $2j = n$ photons in a MZ interferometer employing photon counting measurements, the optimal scaling of precision with photon number is $\delta \phi_e = 1/n$. This Heisenberg limit is \emph{uniquely} achieved by an input
\begin{equation}
|\psi_0 \rangle \mapsto (|j,+j\rangle_y + e^{i \zeta} |j,-j\rangle_y )/\sqrt{2} \: ,
\end{equation}
which takes the form of a `NOON' state \cite{NOON-both} after the first beam-splitter. Traversing the phase element $\theta$ and second beamsplitter, the state emerges with distribution:
\begin{equation}
 p_m(\theta)  \! = \! \frac{(2 j)! \left(1 \! + (-1)^{j+m} \cos \{ 2 j (\theta +\pi/2) - \zeta \}\right)}{4^j(j-m)! (j+m)!} \label{NOON-dist}
\end{equation}
Calculation of Eq.\eqref{F1} gives $\mathcal{J}(\theta) = 4j^2 =  n^2$, independent of $\theta$. This is an exact result, without additional assumptions or approximations. So in the context of local distinguishability this NOON state is indeed optimal, and it has the smallest lower bound on mean-squared phase error. A caveat is that this result is non-constructive: no estimation technique is proposed that might reach the bound \cite{note-on-number-of-trials}.  And despite this local optimality, the NOON distribution has a periodicity $\pi/j$ or $2\pi/n$, causing the distinguishability to saturate quickly for $\Delta > \pi/j$. Thus, NOON states are inappropriate given poor prior knowledge about the phase. But in terms of phase stability, they are the most sensitive to changes, e.g. in tracking a moving target phase \emph{after} its initial acquisition, provided the target is moving slowly enough.

Let us examine other probe states proposed for interferometry. For $| \psi_0 \rangle \mapsto |j,m\rangle_z$, one can use $2 \langle \hat{J}^2_{y}\rangle = \langle \hat{J}^2 \rangle - \langle \hat{J}_z^2 \rangle$, and $\phi_m = 0$ to show in this case $\mathcal{J}(\theta)= 2 [j(j+1)- m^2]$, independent of angle \cite{compare}. Within this family of states the $m=0$ state \cite{Holland-Burnett} gives the best local distinguishability and precision limit via the Cram\'{e}r-Rao bound, $\delta \phi_e \geq [2j(j+1)]^{-1/2}$; lower than the bound found in \cite{Fisher-Optimal-measurements} for the same state and so-called optimal phase measurements. The state is one of equal photon numbers injected at both input modes to the interferometer, $|n/2 \rangle_a \otimes |n/2 \rangle_b $, representing the $n$ photon component of the two-mode squeezed state: $|\lambda \rangle \propto \sum_{\nu=0}^{\infty}\lambda^{n} |\nu \rangle_{a} \otimes | \nu \rangle_{b}$. In contrast, the input state $|j,+j \rangle_z$ returns the `standard' limit, $\delta \phi_e \geq 1 / \sqrt{n}$ for the Cram\'{e}r-Rao bound. This is identical to the result for  $n$ independent experiments carried out with a single photon each. It is also the limit for any single mode state (e.g. coherent light) combined with the vacuum at the MZ input ports, $|\psi\rangle_a \otimes | 0 \rangle_b$. This is expected because $|n\rangle_a \otimes |0\rangle_b \mapsto |j, +j\rangle_z$ is the $n$-photon (or spin-$j$) component of such an input state.

\begin{figure} 
\includegraphics[width=3.2in]{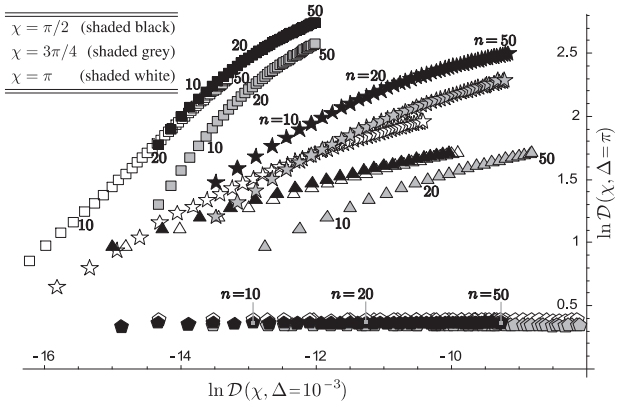}
\caption{\small{Comparing distinguishability $\mathcal{D}(\chi, \Delta)$ in relatively local ($\Delta = 10^{-3}$) and global ($\Delta = \pi$) regimes for various inputs $| \psi_0\rangle$. Each scatter point corresponds to a photon number $2j=n \in [5, 50]$, shaded according to $\chi \in \{ \frac{\pi}{2},\frac{3 \pi}{4},\pi\}$. Pentagons depict NOON states ($\zeta = 0$), triangles $|j,0\rangle_z$, squares $|j, +j\rangle_z$, and stars denote phase states of Eq.\eqref{phase-state} with $\gamma = \pi/2$. NOON states exhibit the best local distinguishability, yet in the $\Delta = \pi$ domain their $\mathcal{D}$ is small and does not improve with photon number. In contrast, $|j, +j\rangle_z$ (corresponding to only one illuminated input port) has strong global characteristics but poor local distinguishability. The phase states (stars) seem to offer the best compromise performance. Notice that the choice of $\chi$ has a significant effect on distinguishability for finite $\Delta$.}} \label{trails}
\end{figure}

Consider the family of phase states \cite{phase-state,Phase-POVM}:
\begin{equation}\label{phase-state}
|j, \gamma \rangle = \frac{1}{\sqrt{2j+1}} \sum_{m=-j}^{j}e^{i m \gamma} |j,m \rangle_y \; ,
\end{equation}
parameterized by real phase $\gamma$. The MZ transformation is $e^{i \theta \hat{J}_y } |j, \gamma \rangle = |j, \gamma + \theta \rangle$. For local distinguishability, Eq.\eqref{Ffinal} gives $\mathcal{J}(\theta) =\frac{4}{3} j (j+1)$, i.e. phase states have the same close-to-optimal scaling as the $|j,0\rangle_z$ input, but with a $2/3$ pre-factor. However, FIG.\ref{trails} illustrates that phase states have much better global distinguishability.

Classical relative entropy is a very useful tool in classifying the fitness of quantum states for phase estimation in both local and global contexts. By interpolating between these two regimes via the a priori phase precision $\Delta$, it has become apparent that a probe state $|\psi_0 \rangle$ displaying strong local distinguishability characteristics may not be optimal in the context of global phase estimation, and vice versa.  Despite this seeming trade-off between local and global properties, some input states have been shown to be `jack of all trades', in particular the phase states (FIG.\ref{trails}). For each probe state the range of $\Delta$ leading to optimal distinguishability, e.g. $\Delta < \pi/j$ for NOON states, may be identified as a new sub-wavelength scale parameter. 

The proportionality of local distinguishability to Fisher information has been emphasized, and we derived an explicit form in terms of a quantum-classical kinetic energy discrepancy. Using this result a proof of the Heisenberg precision bound was given, involving no limiting assumptions of small phase or large photon number. 

For phase estimation protocols with specific global and local distinguishability requirements, the framework we have presented is a flexible and powerful tool in finding an optimal trade-off between sensitivity and photon resources. As our approach is based on the properties of measurement probabilities it is easily extended to other phase detection technologies beyond the prototypical case we consider. Future work should extend the current analysis to a realistic setting incorporating effects of dephasing, thermal noise and, most importantly, photon losses.

G.A.D.'s contribution was made at the Jet Propulsion Laboratory, while holding a Post-doctoral Fellowship of the National
Aeronautics and Space Administration.  J.P.D. acknowledges support from the Army Research Office and the Disruptive Technologies Office. G.A.D. thanks Shunlong Luo and Colin Williams for useful discussions.

\bibliographystyle{apsrev}

\begin{thebibliography}{31}
\bibitem{note-on-number-of-trials}
\bibinfo{note}{For a set of experimental trials $k \sim  \infty$ the maximum likelihood estimation technique is unbiased and approaches the Cram\'{e}r-Rao bound. However, for an $n$-photon input state where $\alpha > 1$ in $\mathcal{J}(\theta)  \propto n^{\alpha}$, the total photon resources $N = k n$ are better utilised in a small number of trials \cite{Braunstein-knee}. Unfortunately, for small $k$ the maximum likelihood estimator is not accurate and unbiased, \cite{MLE-vs-Bayes}. Also, quantum phase distributions can feature multiple equivalent maxima, an unresolvable ambiguity.}
\bibitem{compare}
\bibinfo{note}{This is inclusive of a result of \cite{Hradil-Fisher} derived for $\theta \sim 0$.}
\end{thebibliography}

\end{document}